\documentclass[a4paper,11pt]{article}
\pdfoutput=1 % if your are submitting a pdflatex (i.e. if you have
\usepackage{jheppub} % for details on the use of the package, please
\usepackage[T1]{fontenc} % if needed
\usepackage{bm}
\usepackage{mathrsfs}
\usepackage{amsbsy}
\usepackage{amssymb}
\usepackage{graphicx}
\usepackage{subfigure}
\usepackage{amsmath}
\usepackage{amsfonts}
\usepackage{xcolor}
\usepackage{makecell}

\newcommand{\mpref}[1]{Figure.\ref{#1}}
\newcommand{\be}{\begin{equation}}
\newcommand{\ee}{\end{equation}}
\newcommand{\bea}{\begin{eqnarray}}
\newcommand{\eea}{\end{eqnarray}}

\title{\boldmath Geometric Constraints via Page Curves: Insights from Island Rule and Quantum Focusing Conjecture}

\author[a]{Ming-Hui Yu,}
\author[a]{Xian-Hui Ge,\note{Corresponding author.}}

\affiliation[a]{Department of Physics, College of Sciences, Shanghai University,\\99 Shangda Road, 200444 Shanghai, China}

\emailAdd{yuminghui@shu.edu.cn}
\emailAdd{gexh@shu.edu.cn}

\abstract{Exploring the inverse problem tied to the Page curve phenomenon and island paradigm, we investigate the geometric conditions underpinning black hole evaporation where information is preserved and islands manifest, giving rise to the characteristic Page curve. Focusing on a broad class of static black hole metrics in asymptotically Minkowski or (anti-)de Sitter spacetimes, we derive a pivotal constraint on the blacken factor $f(r)$ for which the island exists and reproduce the Page curve. Specifically, we reveal that a sufficient yet not universally necessary criterion -- manifested in the negativity of the second derivative of $f(r)$, i.e. $f^{\prime \prime} (r)<0$, in proximity to the event horizon where $r \sim r_h+ {\cal O} (G_N)$, ensures the emergence of Page curves in a manner transcending specific theoretical models. This pivotal finding, supported by the tenets of the quantum focusing conjecture.}

\begin{document}
\maketitle
\flushbottom

\section{Introduction} \label{intro}
\quad Black holes are the strongest evidence of general relativity (GR). In modern physics, this surprising and fascinating object has becomes one of the most controversial areas of theoretical physics, When some of the result of quantum mechanics (QM) are inserted into the framework of GR, something amazing occurs. This approach was first proposed by Hawking in 1975 (known as the Hawking radiation) \cite{HR}. However, it leads to a very acute dilemma: the information (loss) paradox \cite{paradox}. QM requires that the evolution of a black hole formed in a pure state must respect the unitary principle, namely, it remains a pure state at the end of evaporation. In contrast, Hawking radiation indicates that radiation in a thermal (mixed) state \footnote{For analytical simplicity, we omit the consideration of the grey-body factor in this study. Consequently, Hawking radiation is effectively modeled as pure black-body radiation, adhering rigorously
to the Planckian spectral distribution.}. It was not until the Page curve was proposed that this issue gradually became sharp \cite{PC1,PC2}. Significant breakthroughs have been made in the last 20 years. A key catalyst was the anti-de Sitter/Conformal Field Theory (AdS/CFT), or the holographic duality \cite{adscft}.
\par The AdS/CFT duality opens a window for us to look at the problem of gravity in AdS from the perspective of CFT though this theory. A milestone work is the RT formula proposed by Ryu and Takayanagi to calculate the holographic entanglement entropy \cite{RT}. The RT formula tells us how to relate the entanglement entropy on the gravitational side by the area of an extremal surface (the RT surface) that is homogeneous with the subregion. Next, the quantum correction of the RT formula is also followed \cite{QRT}. In 2015, the modified RT formula with high-order corrections, the quantum extremal surface (QES) prescription was proposed \cite{QES}.
\par At now, all the problems of evaluating the entanglement entropy at the boundary translate into finding the minimal extremal surface in bulk spacetime. After the Page time, we have another additional extremal surface, which is located inside the event horzion of the evaporating AdS black hole, called the ``island'' \cite{bulk entropy,island rule,entanglement wedge}. Considering its contribution leads to the unitary Page curve. At this point, the black hole information paradox is declared to be preliminarily solved. Interested readers can refer to a nice pedagogical review \cite{review}.
\par The formula for calculating the fine-grained (entanglement) entropy (or the von Neumann entropy) of Hawking radiation obtained by the QES prescription is summarized as the ``island formula'' :
\begin{equation}
S_{\text{Rad}}=\text{Min} \bigg \{ \text{Ext} \bigg[ \frac{\text{Area} (\partial I)}{4G_N} +S_{\text{bulk}} (R \cup I) \bigg] \bigg \}, \label{island formula}
\end{equation}
where $I$ refer to the island region and its boundary is denoted as $\partial I$. The entropy of bulk fields consists of two contributions, namely, the island I inside the black hole and the radiation region $R$ outside the black hole. The words ``Min'' and ``Ext'' guide us to extremize the generalized entropy first to find saddle points,
\begin{equation}
\frac{\partial S_{\text{gen}}} {\partial x^{\mu}} \equiv \frac{\partial} {\partial x^{\mu}} \bigg (\frac{\text{Area} (\partial I)}{4G_N} +S_{\text{bulk}} (R \cup I) \bigg) =0.
\end{equation}
These saddle points correspond to the candidate ``QES''. Then we pick the one with the smallest value, which is the final correct result of the fine-grained entropy of Hawking radiation. In addition, the island formula \eqref{island formula} can be derived equivalently by strict gravitational path integral \cite{replica1,review}:
\begin{equation}
S_{\text{Rad}}= \lim\limits_{n \to 1} \frac{1}{1-n} \log \text{Tr} (\rho_R^n), \label{path integral}
\end{equation}
in which, the contribution of the connected replica wormhole (saddle) will dominate at late times, and the Page curves can be reproduced naturally\footnote{More precisely, all QES configurations are saddle points in the path integral of the replica geometry. The entanglement entropy is minimized to achieve the minimum partition functions. So the entanglement entropy is approximately the minimum entanglement entropy at the saddle point.}.
\par Recently,  studies have demonstrated that the island formula does not depend on the AdS/CFT correspondence and has been applied far beyond the asymptotically AdS black holes. One can refer to a non-exhaustive list of progress in this field \cite{eternalbh,jt1,jt2,2d1,2d2,2d3,2d4,2d5,2d6,2d7,2d8,2d9,high1,high2,high3,high4,high5,high6,high7,high8,high9,high10,high11,high12,high13,high14,high15,high16,high17,high18,high19,high20,high21,high22,high23,high24,high25,high26,high27,high28,high29,high30,high31,high32,high33,high34,high35,high36,high37,high38,high39,high40,high41,high42,high43,high44,high45,high46,high47,high48,high49,high50,high51,high52,high53,high54,high55,high56,high57}.
\par Up to now, most studies have focused on the reproduction of Page curves in \emph{special} spacetime. They all found that islands emerges at late times could curb the growth of entropy and respect the unitarity\footnote{For two-dimensional Liouville black holes, there are no island at late times and the information paradox may not be solved using the island paradigm \cite{2d5}.}. A natural question is what are the constraints on obtaining a unitary Page curve using the island paradigm for \emph{general} spacetime? Or equivalently, we can consider the inverse of this problem: If a Page curve already exists, namely, the unitary is maintained, what constraints does the spacetime geometry need to satisfy? On the other hand, the quantum focusing conjecture (QFC) also has a constraint on the generalized entropy at late times. How does this constraint relate to those imposed by the island paradigm? Therefore, based on QFC perspective, we again consider the requirements of the recurrence of Page curves on spacetime geometry. Incorporating these dual considerations, we discern that the satisfaction of  condition $f^{\prime \prime}(r)<0$ by the second derivative of the blacken function in the vicinity of the horizon, while being a potent yet not universally mandatory requirement, invariably precedes the manifestation of the Page curve. This discovery culminates in the formulation of overarching geometric principles.
\par We begin with a general metric that represents a static black hole. In the static coordinate system under the Schwarzschild gauge, such metrics are written as:
\begin{equation}
ds^2=-f(r)dt^2 +f^{-1} (r)dr^2 +R(r) h_{ij} dx^i dx^j, \label{metric1}
\end{equation}
where the spacetime coordinates are denote as $x^{\mu}=(t,r,x^{i}),(i=1,2...)$. The horizon metric $h_{ij}$ is a function of $x^{i}$ only. To guarantee the existence of a black hole solution, we need to impose some requirements on the blacken function $f(r)$: It must have simple and positive zeros, and then it is also required to have a value for its corresponding radial coordinates that exceeds the horizon and extends to the space-like infinity. Only in this way is the domain of exterior communication is ``outside'' the black hole.
\par In some special cases, the blacken functions for radial and time coordinates are not equal, and the function $R(r)$ has a complicated form\footnote{Here, we did not consider the special hairy black hole for convenience. Their metrics can not be written in the Schwarzschild form \cite{hairy}. In addition, since dynamic black holes have the back-reaction on spacetime, we also ignore the evaporating black hole model and focus mainly on eternal black holes in this paper.}. Actually, this corresponds to the configuration with the Einstein-Maxwell-dilation field equation\footnote{For instance, for Garfinkle-Horowitz-Strominger black holes, the metric cannot be written in the form of \eqref{metric1} \cite{high31}; for Kaluza-Klein black holes, $R(r)$ is a function of the dilation field $\phi$ \cite{high15}.} \cite{dilaton}. For convenience, we ignore these few special examples in the paper and assume that black hole solutions can all be written in the form \eqref{metric1}. Moveover, when the cosmological constant $\Lambda$ is non-positive, it is asymptotically associated with Minkowski or AdS black holes. They usually have only one horizon, except for topological black holes. However, for a positive cosmological constant, there were black holes with multiple horizons. For simplicity, we focus mainly on the case of a single horizon. In the case of multiple horizons, the corresponding calculation only requires parameter substitution without affecting the physical meaning. One can refer to \cite{high37} for the explicit calculations.
\par The rest of the paper is organized as follows. In section \ref{island paradigm}, we calculate the entanglement entropy for Hawking radiation by the island paradigm. We first prove that island is absent at early times. Subsequently, we focus on the behavior of entropy at late times. We derive the constraint condition that the spacetime geometry needs to satisfy when the island appears, and we must obtain a unitary Page curve. In section \ref{QFC}, we apply the QFC to test our result and acquire a self-consistent conclusion. Finally, we display the discussions and summary in section \ref{conclusion}. The Planck units $\hbar=k_B=c=1$ is used through the paper.

\section{Island Paradigm for Black Holes} \label{island paradigm}
\quad In this section, we evaluate the entanglement entropy of Hawking radiation using the island formula \eqref{island formula}. We directly assume that there is an island in black hole spacetime due to the fact that islands are necessary and sufficient to reproduce the Page curve based on the island paradigm. We investigate the behavior of the generalized entropy in the early and late stage respectively. Consequently, we indicate that there no islands at early times and leads to information loss. Then, we focus on the behavior at late times. Finally, We obtain a constraint equation for the spacetime geometry to ensure the appearance of Page curves.
\par A schematic of the Penrose diagram is shown in \mpref{penrose} \footnote{For the bath region, it refer to half-Minkowski spacetime. We usually assume that bath regions have no gravitational effect, or that the gravitational effect can be ignored. Some studies have considered the gravitational bath \cite{high24}.}. In order to extend the metric \eqref{metric1} to the left and right wedges, a Kruskal transformation is allowed:
\begin{figure}[htb]
\centering
\subfigure[\scriptsize{}]{\label{penrose1}
\includegraphics[scale=0.25]{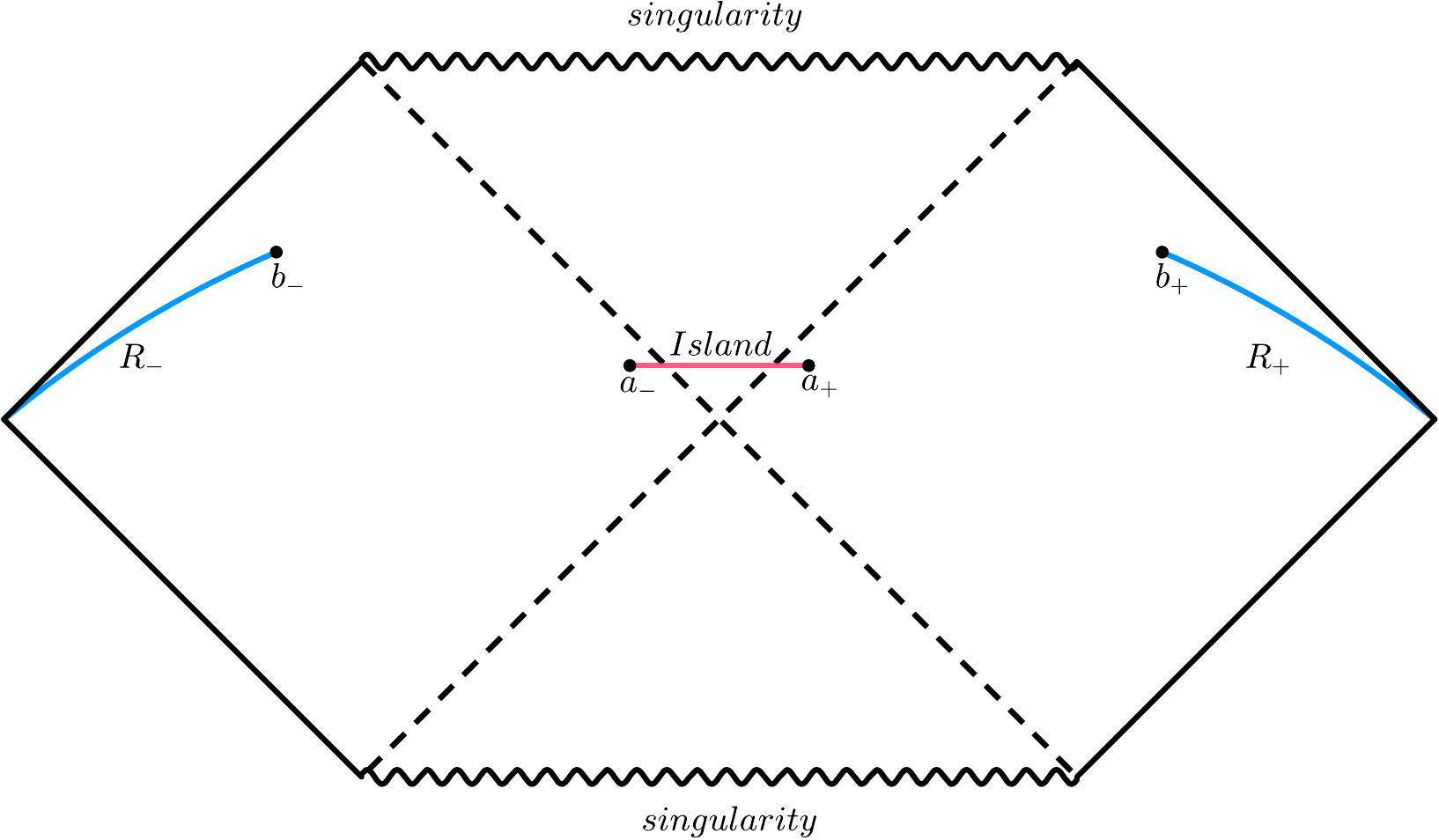}
}
\quad
\subfigure[\scriptsize{}]{\label{penrose2}
\includegraphics[scale=0.25]{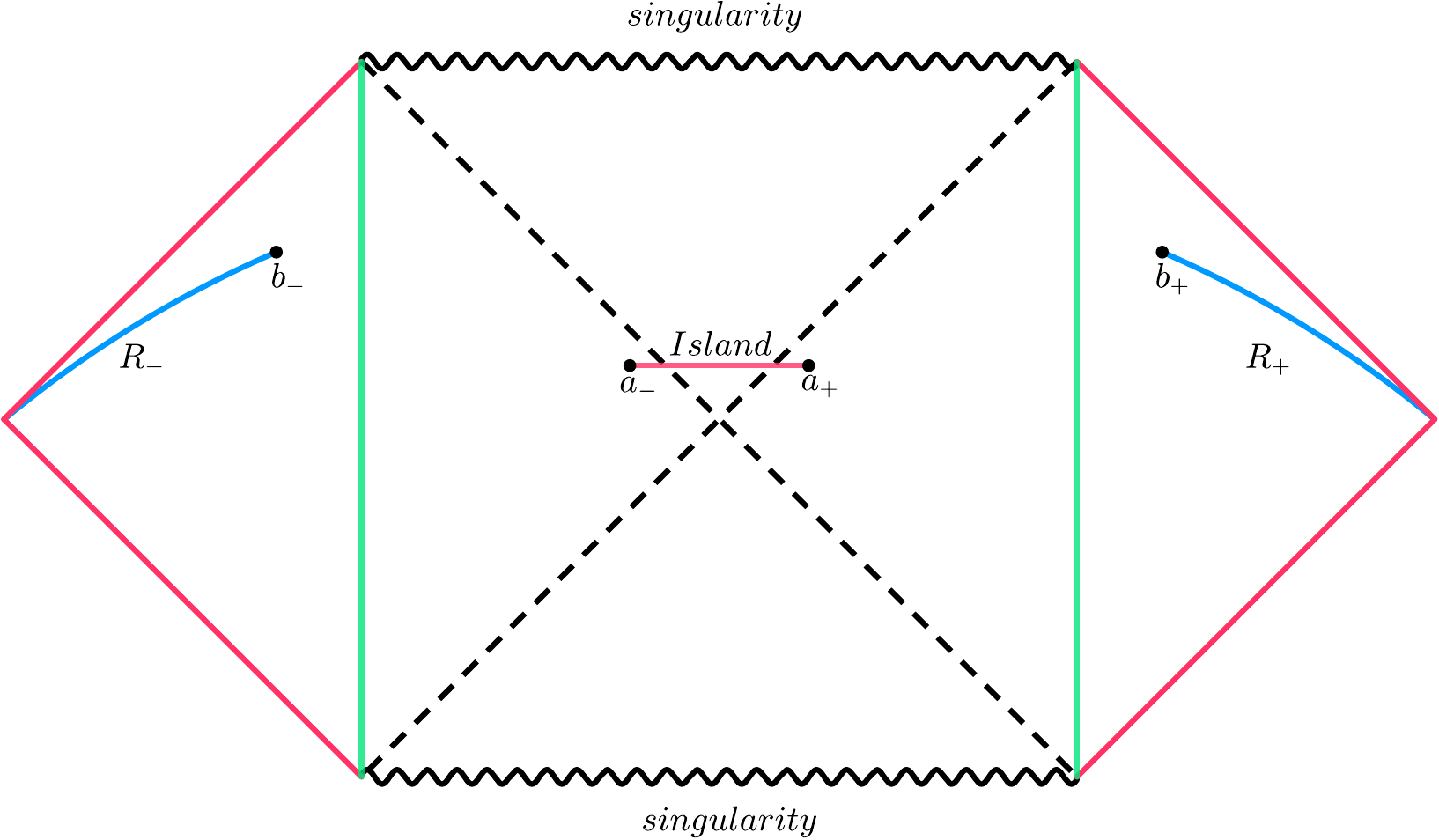}
}
\caption{The schematic Penrose diagram of black holes (with the single horizon). The radiation regions are denoted by $R{\pm}$, and their boundaries are the cut-off surfaces. The coordinates of the radiation are $b_{\pm}=(\pm t_b,b)$. The coordinates of the islands boundaries are $a_{\pm}=(\pm t_a,a)$. On the left, this represents an asymptotically flat black hole. Hawking radiation can naturally diffuse to null infinity. On the right, it represents an asymptotically AdS black hole in thermal equilibrium with the bath (red region). We then impose the transport boundary condition on the black hole region (black region) \cite{island rule}. In such way, Hawking radiation can also be collected by observers at space-like infinity.}
\label{penrose}
\end{figure}
\begin{equation}
\begin{split}
\text{Right Wedge}: \ \ &U \equiv -e^{-\kappa u}=-e^{\kappa (t-r_{\star})}; \qquad V \equiv +e^{+ \kappa v}=+e^{+\kappa (t+r_{\star})}, \\
\text{Left Wedge}: \ \  &U \equiv +e^{-\kappa u}=+e^{\kappa (t-r_{\star})}; \qquad V \equiv -e^{+ \kappa v}=-e^{+\kappa (t+r_{\star})}, \label{kruskal}
\end{split}
\end{equation}
with the surface gravity $\kappa$:
\begin{equation}
\kappa \equiv 2\pi T_H = \frac{1}{2} f^{\prime}(r_h), \label{surface gravity}
\end{equation}
where $T_H$ is the Hawking temperature and $\prime$ represents the derivative with respect to the radial coordinate $r$, and $r_h$ is denoted the radius of event horizons. Here we set $f(r_h)=0$. The tortoise coordinates is defined by
\begin{equation}
r_{\star} (r) = \int^r \frac{1}{f(r)} d\tilde{r}. \label{tortoise}
\end{equation}
After the Kruskal transformation, the metric \eqref{metric1} can be recast as:
\begin{equation}
ds^2=-\Omega^2(r) dU dV +R(r) h_{ij} dx^i dx^j, \label{kruskal tranformation}
\end{equation}
with the conformal factor\footnote{We assume that the bath region is a Minkowski patch without gravitational effect. So, for the bath region, we have $f(r)=1, r_{\star}=r$ and can obtain the expression \eqref{conformal bath} from \eqref{conformal bh}.}:
\begin{subequations}
\begin{align}
&\Omega_{\text{BH}}(r)= \frac{\sqrt{f(r)}}{\kappa e^{\kappa r_{\star}(r)}}, \label{conformal bh} \\
&\Omega_{\text{bath}}(r)=\frac{1}{\kappa e^{\kappa r}}.  \label{conformal bath}
\end{align}
\end{subequations}

\subsection{No islands at early times}
\quad At first, we consider the issue from the perspective of an observer at the infinity, which means that $b-r_h \gg r_h$. In this case, we can only consider two-dimensional (2D) massless scalar field and assume the ``s-wave approximation'' is hold\footnote{Although there exists the massive modes in Kaluza-Klein tower of the spherical part, only the s-wave without non-zero angular momentum has contribution when the distance is much larger than the coherence length of massive modes.}. In addition, we assume that the black holes is a pure state at initial time.
\par In the construction of the no-island, only radiation remains. We can only consider the complementary region of radiation based on the complementary of von Neumann entropy. As a consequence (see Appendix \ref{appendixa}):
\begin{equation}
\begin{split}
S_{\text{Rad}}&=S(R)=\frac{c}{6} \log \Big[d^2(b_-,b_+) \Omega(b_-) \Omega(b_+)  \Big] \\
              &= \left \{
              \begin{array}{lr} \frac{c}{6} \log \bigg(\frac{4f(b)}{\kappa^2} \cosh^2 (\kappa t_b)\bigg), \qquad \text{for asymptotically flat black holes}  &\\
              \frac{c}{6} \log \bigg( \frac{4}{\kappa^2} \cosh^2(\kappa t_b) \bigg), \qquad \ \ \ \text{for asymptotically AdS black holes}&
              \end{array} \right. \label{ee without1}
\end{split}
\end{equation}
where $c$ represents the central charge. In the limit of late times and large distances, we can take the approximation: $\cosh(\kappa t_b) \simeq \frac{1}{2} e^{\kappa t_b}$. Then the above equation is equal to:
\begin{equation}
S_{\text{Rad}}(\text{without island}) \simeq \frac{c}{3} \kappa t_b. \label{result1}
\end{equation}
Apparently, without island construction, the entanglement entropy of radiation grows linearly with time at late times, which leads to the information loss and consistent with Hawking's view. In addition, the result \eqref{result1} does not depend on the geometry $f(r)$, which implies that the information paradox is always exists.
\par Next, we turn to the construction with an island to obtain the Page curve. Similarly, referring to the Penrose diagram in \mpref{penrose}, we see the entire Cauchy slice is divided into three intervals. For the disconnected union interval $R  \cup  I$, the expression of the entanglement entropy is converted from \eqref{ee without1} (only valid for a single interval) to the following form \cite{ee formula1,ee formula2}:
\begin{equation}
\begin{split}
S_{\text{bulk}} (R \cup I) &=\frac{c}{3} \log \bigg(  \frac{d(a_+,a_-) d(b_+,b_-) d(a_+,b_+) d(a_-,b_-)}{d(a_+,b_-)d(a_-,b_+)}     \bigg) \\
                           &=\frac{c}{6} \log \Big[16 \Omega^2(a) \Omega^2(b) e^{2\kappa (r_{\star}(a)+r_{\star}(b))}  \cosh^2(\kappa t_a) \cosh^2 (\kappa t_b)  \Big] \\
                           &+\frac{c}{3} \log \bigg[ \frac{\cosh[\kappa (r_{\star}(a)-r_{\star}(b))] - \cosh[\kappa (t_a-t_b)] }{\cosh[\kappa (r_{\star}(a)-r_{\star}(b))] + \cosh[\kappa(t_a+t_b)]}    \bigg], \label{ee with1}
\end{split}
\end{equation}
where
\begin{subequations}
\begin{align}
&\Omega^2(a) \Omega^2(b) = \Omega^2_{\text{BH}} (a) \Omega^2_{\text{BH}}(b)=\frac{f(a)f(b)}{\kappa^4 e^{2\kappa (r_{\star}(a)+ r_{\star}(b))}}, \ \ \  \text{for asymptotically flat cases}  \\
&\Omega^2(a) \Omega^2(b) = \Omega^2_{\text{BH}} (a) \Omega^2_{\text{bath}}(b)=\frac{f(a)}{\kappa^4 e^{2\kappa (r_{\star}(a)+b) }}.\qquad \text{for asymptotically AdS cases}
\end{align}
\end{subequations}
Accordingly, the generalized entropy read as\footnote{Hereafter, we only present the results for asymptotically flat black holes for the sake of simplicity. In order to fit the AdS black holes, one simply set $f(b)=1$ and $r_{\star}(b)=b$.}:
\begin{equation}
\begin{split}
S_{\text{gen}} &=\frac{A(a)}{2G_N} +\frac{c}{6} \log \bigg[ \frac{16 f(a)f(b)}{\kappa^4} \cosh^2(\kappa t_a) \cosh^2(\kappa t_b)  \bigg] \\
               &+\frac{c}{3} \log \bigg[ \frac{\cosh[\kappa (r_{\star}(a)-r_{\star}(b))] - \cosh[\kappa (t_a-t_b)] }{\cosh[\kappa (r_{\star}(a)-r_{\star}(b))] + \cosh[\kappa(t_a+t_b)]}   \bigg],  \label{generalized entropy1}
\end{split}
\end{equation}
where $A(a)$ is the area of island, which is a positive constant and determined by the function $R(r)$ \eqref{metric1}.
\par At very early times, we assume that $t_a \simeq t_b \simeq 0 \ll \kappa b$. Then the generalized entropy becomes:
\begin{equation}
S_{\text{gen}}^{(\text{early})} \simeq \frac{A(a)}{2G_N} +\frac{c}{6} \log \bigg[ \frac{16 f(a)f(b)}{\kappa^4} \cosh^2 (\kappa t_a) \cosh^2 (\kappa t_b) \bigg]. \label{early entropy}
\end{equation}
In order to obtain the QES, we extremize the above expression with respect to $a$ and $t_a$:
\begin{equation}
\frac{\partial S_{\text{gen}}^{(\text{early})}}{\partial t_a}=\frac{c \kappa}{3} \tanh (\kappa t_a)=0. \label{wrt t}
\end{equation}
The only solution is $t_a=0$, so the approximation is right. Then the location of QES can be obtained by the following equation:
\begin{equation}
\frac{\partial S_{\text{gen}}^{\text{(early)}}}{\partial a}= \frac{A^{\prime}(a)}{2G_N} +\frac{c}{6} \frac{f^{\prime} (a)}{f(a)}=0. \label{wrt a}
\end{equation}
We can rewrite this expression to obtain the constraint equation that is satisfied if the island appears at early times:
\begin{equation}
- \frac{3 A^{\prime}(a) f(a)}{f^{\prime}(a)} =cG_N \ll 1. \label{constraint1}
\end{equation}
Here, we assume that the central charge is relatively small: $0<cG_N \ll 1$. This is because the area term $A^{\prime} (a)$ is finite. The only possible solution to the above equation is\footnote{When $a>r_h$, $f(a)>0$ and its derivative function $f^{\prime}(a)$ can never be negative due to the fact that $f^{\prime}(r_h)$ is a function related to the surface gravity $\kappa = \frac{f^{\prime}(r_h)}{2}$>0 \eqref{surface gravity}. Thus the only case is that $a<r_h$.}:
\begin{equation}
f^{\prime}(a)<0, \qquad a\gtrsim 0.
\end{equation}
However, the size of island is now less than the Planck length $\ell_p (\sim \sqrt{G_N}$ for 4D case). One should discard this solution, which is consistent with our expectation. Namely, we can infer that islands absent at early times, which does not depend on the metric \eqref{metric1}.

\subsection{Constraints on the background geometry at late times}
\quad By contrast, at large distances and late times, the left wedge and right wedges are significantly separated. To simplify this, we can perform the following approximation \cite{high4}:
\begin{equation}
d(a_+,a_-) \simeq d(b_+,b_-) \simeq d(a_+,b_-) \simeq d(a_-,b_+) \gg d(a_+,b_+) \simeq d(a_-,b_-). \label{ope}
\end{equation}
Then, the entanglement entropy at late times is simplified as:
\begin{equation}
\begin{split}
S_{\text{gen}}^{(\text{late})} &\simeq \frac{A(a)}{2G_N}+\frac{c}{3} \log [d(a_+,b_+)d(a_-,b_-)] \\
                               &=\frac{A(a)}{2G_N} +\frac{c}{6} \log \bigg[ \frac{4f(a)f(b)}{\kappa^4}  \bigg( \cosh(\kappa (r_{\star}(a)-r_{\star}(b)))-\cosh(\kappa(t_a-t_b)) \bigg)^2 \bigg]. \label{late entropy}
\end{split}
\end{equation}
In same way, we extremize it with respect to time $t_a$ firstly,
\begin{equation}
\frac{\partial S_{\text{gen}}^{\text{(late)}} }{\partial t_a}= -\frac{c}{3} \frac{\kappa \sinh[\kappa (t_a-t_b)]}{\cosh[\kappa (r_{\star}(a) -r_{\star}(b))] - \cosh[\kappa (t_a-t_b)]} =0. \label{wrt t2}
\end{equation}
The only solution is to set $t_a$ equal to $t_b$, and then substitute this relation into the original expression with respect to $a$,
\begin{equation}
\frac{\partial S_{\text{gen}}^{(\text{late})} }{\partial a}=\frac{A^{\prime} (a)}{2G_N} +\frac{c}{6} \bigg[ \frac{f^{\prime}(a)}{f(a)} + \frac{2\kappa}{f(a)} \coth \Big[\frac{\kappa}{2}(r_{\star}(a) -r_{\star}(b))\Big] \bigg]=0, \label{wrt a2}
\end{equation}
where we have used $r_{\star}^{\prime}(a)=\frac{1}{f(a)}$. Following \eqref{constraint1}, we rewrite this expression as:
\begin{equation}
\frac{3 A^{\prime}(a) f(a)  }{2\kappa  \frac{e^{\kappa x}+1}{e^{\kappa x}-1} -f^{\prime}(a) } =cG_N \ll 1, \label{constraint2}
\end{equation}
where $x\equiv r_{\star}(b) - r_{\star}(a) \gg r_h$. Now we make the near horizon limit: $a\simeq r_h$ and obtain:
\begin{subequations}
\begin{align}
&f(r) \simeq f^{\prime}(r_h)(r-r_h) + {\cal O} [(r-r_h)^2],  \label{approximation1}\\
&r_{\star}(r)=\int^r \frac{d \tilde{r}}{f(r)} \simeq \frac{1}{2\kappa} \log \bigg| \frac{r-r_h}{r_h}  \bigg|. \label{approximation2}
\end{align}
\end{subequations}
Substituting these equations into \eqref{constraint2}, yields the following constraint equation:
\begin{equation}
0<y(r)=\frac{3 A^{\prime}(a) f^{\prime}(r_h) (a-r_h)^2 }{f(a) \frac{e^{\kappa x}+1}{e^{\kappa x}-1} -(a-r_h) f^{\prime}(a)} \simeq \frac{3 A^{\prime}(a) f^{\prime}(r_h) (a-r_h)^2}{f^{\prime}(a) \frac{e^{\kappa x}+1}{e^{\kappa x}-1} -f^{\prime \prime}(a)} =cG_N \ll 1. \label{constraint3}
\end{equation}
Firstly, we find that there is a solution to this equation in almost all cases, with the exception of the extremal case (see Appendix \ref{appendixb}). Additionally, it should be note that the island is consistently located outside of the event horizon:
\begin{equation}
a=r_h+ \frac{cG_N}{6 A^{\prime}(r_h)} e^{\kappa r_{\star}(b)} +\frac{c^2 G_N^2}{72 A^{\prime 2}(r_h) r_h} +{\cal O} [(cG_N)^3]. \label{location}
\end{equation}
Therefore, according to this location, we can obtain the entanglement entropy of radiation at late times:
\begin{equation}
\begin{split}
S_{\text{Rad}} &= \frac{A(r_h)}{2G_N} +{\cal O} (G_N) \\
        &\simeq 2S_{\text{BH}}.  \label{result2}
\end{split}
\end{equation}
It is what we expect. Recall the result without island \eqref{result1}, the Page time is determined by
\begin{equation}
t_{\text{Page}}=\frac{6S_{\text{BH}}}{c \kappa} =\frac{3 S_{\text{BH}}}{c \pi T_H}. \label{page time}
\end{equation}
Besides, we can also calculate the scrambling time as a by-product. Drawing from the insights of the Hayden-Preskill thought experiment \cite{HP}, it is posited that an external observer, situated asymptotically relative to the black hole, must patiently await the elapsed duration known as the "scrambling time" before information initially engulfed by the black hole can be retrieved through analyzing the emitted Hawking radiation. In the language of the entanglement wedge reconstruction, the scrambling time corresponds to the time when the information reaches the boundary of island ($r=a$) from the cut-off surface ($r=b$) \cite{entanglement wedge}:
\begin{equation}
\begin{split}
t_{\text{scr}} &\equiv \text{Min} [v(t_b,b)-v(t_a,a)] =r_{\star}(b) -r_{\star}(a) \\
               &\simeq r_{\star}(b) - \frac{1}{2 \kappa} \log \frac{a-r_h}{r_h} \simeq \frac{1}{2\kappa} \log \frac{A^{\prime}(r_h)r_h}{cG_N} \\
               &\simeq \frac{1}{2\pi T_H} \log S_{\text{BH}}, \label{scrambling time}
\end{split}
\end{equation}
where $t_a, t_b$ is the time of sending and receiving information, respectively. In the penultimate line, we employed the approximation delineated in equation \eqref{approximation2} to facilitate our calculations. Concludingly, we adopted the established outcome for the four-dimensional scenario, aligning seamlessly with the findings reported in the seminal Hayden-Preskill thought experiment \cite{scrambling,acoustic}, thus ensuring theoretical consistency.
\par Above all, we protects the unitary by the island formula. In particular, by combining the discussion above with the expression \eqref{constraint3}, we also obtain a \emph{sufficient but not necessary condition} for deriving the Page curve:
\begin{equation}
f^{\prime \prime}(r_h+ {\cal O}(G_N) )<0, \label{condition1}
\end{equation}
Specifically, the radial coordinate $r$  is confined to a region situated just outside the event horizon, adhering to the condition $r \gtrsim r_h$, reflecting our focus on the immediate vicinity of the horizon through implementation of the near-horizon approximation. Since we make the near horizon limit. In other words, if the above constraint is true, the equation \eqref{constraint3} must have a solution, and then there is an island, which makes the entanglement entropy satisfy the Page curve. We present the results of calculations for some typical black holes in Appendix \ref{appendixc}. The impact of the condition on the result will be discussed in detail in the following section.

\section{Island and Quantum Focusing Conjecture} \label{QFC}
\quad Up to the present, we calculate the Page curve using the island formula \eqref{island formula}. Combing the results of the previous section, we obtain the behavior of entanglement entropy in the entire process of black hole evaporation is
\begin{equation}
S_{\text{Rad}}=\text{Min} \Bigg[\frac{2 \pi c}{3} T_H t, 2S_{\text{BH}}\Bigg]. \label{result3}
\end{equation}
In particular, we find that if the constraint condition \eqref{constraint3} is satisfied, the Page curve must be reproduced, and there must exist an island outside the event horizon \eqref{location}. This conclusion is universal and not depend on the explicit form of the metric \eqref{metric1}. In this sense, we provide the constraint conditions of spacetime when Page curve is established.
\par Now in this section, we further prove the correctness of the constraint from the perspective of QFC. The classical focusing theorem asserts that the expansion $\theta$ of the congruence of null geodesic never increases:
\begin{equation}
\frac{d\theta}{d \lambda} \le 0, \label{cfc}
\end{equation}
where $\lambda$ is the affine parameter. An important application of this theorem is to prove the second law of black holes. For a black hole with area $A$ and entropy $S_{\text{BH}}$($=\frac{A}{4G_N}$), the expansion $\theta$ is defined by:
\begin{equation}
\theta=\frac{1}{A} \frac{dA}{d\lambda}. \label{expansion1}
\end{equation}
Then one can infer to the second law: $dS_{\text{BH}} \ge0$.
\par However, once quantum effects are considered, i.e., the black hole emits Hawking radiation. The second law is violated. For the sake of rationality, this law should be upgraded to the generalized second law. Accordingly, the black hole entropy should be replaced by the generalized entropy: $dS_{\text{gen}} \ge0$. Therefore, the classical focusing theorem is also being extended to the QFC \cite{qfc1,qfc2}, in which the quantum expansion is given by replacing the area in the classical expansion with the generalized entropy \eqref{late entropy}:
\begin{equation}
\frac{d \Theta}{d \lambda} \le 0, \label{qfc1}
\end{equation}
where $\Theta$ is the quantum expansion, which can be expressed in terms of the generalized entropy:
\begin{equation}
\Theta = \frac{1}{A} \frac{d}{d\lambda} S_{\text{gen}}. \label{expansion2}
\end{equation}
\par Now, we investigate the QFC for the construction with an island. For the entanglement entropy at late times \eqref{late entropy}, the quantum expansion is written as:
\begin{equation}
\begin{split}
\Theta&=\frac{1}{A} \frac{d}{d \lambda} S_{\text{gen}}\\
      &=\frac{1}{A} \frac{dv_b}{d\lambda} \bigg[ \frac{\partial S_{\text{gen}}}{\partial v_b}  +\frac{dv_a}{dv_b} \frac{\partial S_{\text{gen}}}{ \partial v_a} +\frac{du_a}{dv_b} \frac{\partial S_{\text{gen}}}{\partial u_a}     \bigg]. \label{expansion3}
\end{split}
\end{equation}
Here we introduce the affine parameter \cite{qfc2}:
\begin{equation}
\lambda \equiv - \frac{\partial r(u,v)}{\partial u} dv, \label{affine}
\end{equation}
for simplicity. Due to the fact that QES makes the entanglement entropy to extremized, which means that:
\begin{equation}
\frac{\partial S_{\text{gen}}}{\partial u_a}=\frac{\partial S_{\text{gen}}}{\partial v_a}=0. \label{qes}
\end{equation}
Then we have
\begin{equation}
\begin{split}
\Theta&=\frac{1}{A} \frac{dv_b}{d \lambda} \frac{\partial S_{\text{gen}}}{\partial v_b} = \frac{\partial S_{\text{gen}}}{\partial t_b} +f(b) \frac{\partial S_{\text{gen}}}{\partial b} \\
&=\frac{f(b) A^{\prime}(b)}{2G_N} - \frac{c \kappa}{3} \coth \bigg[\frac{\kappa}{2}((t_a-t_b)+(r_{\star}(a) - r_{\star}(b)))\bigg] + \frac{c}{6} f^{\prime}(b) >0. \label{expansion4}
\end{split}
\end{equation}
Therefore, the entanglement entropy is always increasing with the null time $v_b$ and the quantum expansion is positive. Moreover, following the QFC, we obtain the derivative of the quantum expansion as:
\begin{equation}
\begin{split}
\frac{d\Theta}{d\lambda} &= \frac{d}{d\lambda} \bigg(\frac{1}{A} \frac{dS_{\text{gen}}}{d\lambda}  \bigg) \\
                         &=\frac{1}{d\lambda} \bigg( \frac{1}{A} \frac{dv_b}{d\lambda} \frac{\partial S_{\text{gen}}}{\partial v_b} \bigg)= \frac{1}{f(b)} \frac{d}{dv_b} X \\
                         &= - \frac{[A^{\prime}(b)]^2-A(b) A^{\prime \prime}(b)}{2G_N A^2(b)} - \frac{c}{6A^2(b) f^2(b)} (Y+Z), \label{qfc2}
\end{split}
\end{equation}
where
\begin{subequations}
\begin{align}
X&= \frac{A^{\prime}(b)}{A(b)G_N} -\frac{c \kappa}{3A(b) f(b)} \coth \bigg[\frac{\kappa}{2}((t_a-t_b)+(r_{\star}(a) - r_{\star}(b)))\bigg] +\frac{c}{6A(b)} \frac{f^{\prime}(b)}{f(b)}, \label{x} \\
Y&=-f(b) A^{\prime}(b) \bigg( 2\kappa  \coth \Big[\frac{\kappa}{2}((t_a-t_b)+(r_{\star}(a) - r_{\star}(b)))\Big] -f^{\prime}(b) \bigg) >0, \label{y} \\
Z&= A(b) \bigg[ \kappa^2 \text{csch}^2 \Big[\frac{\kappa}{2}((t_a-t_b)+(r_{\star}(a) - r_{\star}(b)))\Big] -2\kappa f^{\prime}(b) \times \notag \\ & \coth \Big[\frac{\kappa}{2}((t_a-t_b)+(r_{\star}(a) - r_{\star}(b)))\Big] + (f^{\prime}(b))^2 -f(b) f^{\prime \prime}(b) \bigg]. \label{z}
\end{align}
\end{subequations}
In above calculations, the QES condition \eqref{qes} is used to be simplified. The first term is related to the area, which is always negative\footnote{Here we do not consider the special 2D case for simplicity, where the ``area'' is dependent on the dilation.}. Therefore, one can obviously see that if the QFC is hold, the only requirement is that $Z>0$. Further, since the first three terms of $Z$ are all positive, the we can find that as long as
\begin{equation}
f^{\prime \prime}(r>r_h)<0, \label{condition2}
\end{equation}
where the radial coordinate $r$ can extended near the cutoff surface $r \sim b \gg r_h$. Then, there must be $\frac{d\Theta}{d\lambda}<0$, i.e. the QFC is satisfied.
\par In summary, we verify our previous conclusions from the perspective of QFC. The derived from QFC result \eqref{condition2} contain the previous result from island paradigm \eqref{condition1}. Namely, the applicability of QFC is wider. Therefore, we can conclude that: A sufficient condition for a Page curve for general spacetime \eqref{metric1} to exist is the second derivative of the blacken function is negative in the near horizon region. We stress that this conclusion is only valid at the semi-classical level, where the back-reaction of black holes on spacetime can be ignored. Now we display some physical meaning of the result. When the ``s-wave approximation'' is allowed, we can only focus on the radial part of the metric \eqref{metric1} and ignore the angular momentum. For the metric \eqref{metric1}, it can be reduced to:
\begin{equation}
ds^2=-f(r)dt^2 +f(r)dr^2. \label{metric2}
\end{equation}
The second derivative of blacken function $f(r)$ is associated to the Ricci scalar curvature $R$, and can be positive or negative:
\begin{equation}
R=-f^{\prime \prime}(r). \label{Ricci}
\end{equation}
 Combine with the result \eqref{condition1}, we can summarize that for the black hole spacetime with a positive Ricci scalar curvature, there always exists islands outside event horizons and Page curves can be derived. Moreover, the QFC is always satisfied in such cases. In brief, we verify the rationality of the result \eqref{condition1} again through the QFC.

\section{Discussion and Conclusion} \label{conclusion}
\quad In conclusion, our study significantly contributes to the comprehension of black hole evaporation dynamics and the resolution of the information paradox, leveraging the insights from the Page curve and island paradigm. By examining diverse static black hole metrics within asymptotically flat or (anti-)de Sitter spacetimes, we revealed a pivotal geometric requirement: A metric with a negative second derivative near the horizon, $f^{\prime \prime} (r)<0$, in proximity to the event horizon where $r \sim r_h+ {\cal O} (G_N)$. which is fundamental for island formation and the consequent derivation of Page curves. This discovery affirms the universality of Page curves, transcending model-specific restrictions and reinforcing the compatibility of information conservation within the semi-classical gravity framework.
\par Our findings harmoniously intertwine quantum entanglement principles with the geometry of spacetime, with the QFC robustly supporting our conclusions. This integration bolsters our understanding of the intricate balance between information conservation, geometric configurations, and evaporative black hole dynamics, thus enriching the quantum gravity discourse.
\par We present a comprehensive method for calculating Page curves using a generalized metric \eqref{metric1}, where the entanglement entropy of radiation follows \eqref{result3}. Notably, islands are positioned outside the event horizon \eqref{location}, and we establish a sufficient condition \eqref{condition1} for island emergence. This methodology sets a benchmark for employing the island paradigm in Page curve computations.
\par Furthermore, we identify a pair of conditions (\eqref{condition1} and \eqref {condition2}) that jointly ensure the reproduction of unitary Page curves and adherence to QFC, thereby outlining a stringent criterion for island existence. Our analysis also implicates Ricci curvature, suggesting that spacetimes with positive Ricci curvature facilitate unitarity maintenance, whereas those with negative Ricci curvature necessitate deeper examination of the blacken function $f(r)$ and the fulfillment of the constraint \eqref{constraint3}.

\begin{acknowledgments}
We would thank to Shuyi Lin, Ruidong Zhu and Xiaokai He for discussions related to island rule. The study was partially supported by NSFC, China (Grant No. 12275166 and No. 12311540141).
\end{acknowledgments}

\appendix
\section{Entanglement Entropy in Curved Spacetime}\label{appendixa}
\quad In this appendix, we briefly give the expression of entanglement entropy in curved black hole background and discuss what should be pay attention to when using them.
\par Initially, different from the 2D simple case, the expression of entanglement entropy in the higher-dimensional scenario is complicated and has an area-like divergent. Namely, the entropy for matter fields has following expression:
\begin{equation}
S_{\text{bulk}} (R \cup I) = \frac{\text{Area}(\partial I)}{\epsilon^2} +S_{\text{bulk}}^{\text{finite}} (R \cup I), \label{matter entropy}
\end{equation}
where $\epsilon$ is the cutoff, which is dominates the area-like divergent term. Then we can absorb this term by renormalizing the Newton constant:
\begin{equation}
\frac{1}{4G_N^{(r)}} \equiv \frac{1}{4G_N}+\frac{1}{\epsilon^2}. \label{newton}
\end{equation}
As consequent, we can replace the corresponding part of island formula \eqref{island formula} with $G_N^{(r)}$ and $S_{\text{bulk}}^{\text{finite}} (R \cup I)$, respectively, to yield a finite contribution of the entanglement entropy. Thus, the entanglement entropy in higher-dimensional spacetime is
\begin{equation}
S_{\text{Rad}}=\text{Min} \bigg \{ \text{Ext} \bigg[ \frac{\text{Area}(\partial I)}{4G_N^{(r)}} + S_{\text{bulk}}^{\text{(finite)}} (R \cup I) \bigg] \bigg \}. \label{island formula2}
\end{equation}
\par Secondly, due to the s-wave approximation, the renormalized von Neumann entropy in vacuum CFT$_2$ in \emph{flat} spacetime $ds^2=-dx^+dx^-$ (with the light cone coordinate $x^{\pm}=t \pm r$) is \cite{ee formula1,ee formula2}
\begin{equation}
S_{\text{bulk}}(A \cup B)=\frac{c}{3} \log (d_{AB}), \label{ee flat}
\end{equation}
with
\begin{equation}
d_{AB}\equiv \sqrt{[x^+(A)-x^+(B)][x^-(B)-x^-(A)]}, \label{geodesic1}
\end{equation}
in the geodesic distance between points $A$ and $B$ in flat metric. In order to apply the formula \eqref{island formula2} to the \emph{curved} spacetime, we need to perform the Wely transformation into curved 2D metric $ds^2_{2D}=-\Omega^2(x^+,x^-) dx^+dx^-$ \cite{bulk entropy}. After the Weyl transformation, we finally obtain the entanglement entropy in general 2D spacetime is \cite{2d2}:
\begin{equation}
S_{\text{bulk}} (A \cup B)=\frac{c}{6} \log \bigg[ d^2(A,B) \Omega(A) \Omega(B)   \bigg] \Bigg|_{t_{\pm}=0}. \label{ee curved}
\end{equation}

\section{Derivation of the Location of Islands} \label{appendixb}
\quad In this appendix, we display the details of the expression \eqref{location}. From the equation \eqref{constraint3}
\begin{equation}
0<y(r)=\frac{3 A^{\prime}(a) f^{\prime}(r_h) (a-r_h)^2 }{f(a) \frac{e^{\kappa x}+1}{e^{\kappa x}-1} -(a-r_h) f^{\prime}(a)}=cG_N \ll 1. \label{b1}
\end{equation}
For this equation to have a solution, the denominator has to be greater than zero, since the numerator is positive, namely,
\begin{equation}
f(a) \cdot \frac{e^{\kappa x}+1}{e^{\kappa x}-1} -(a-r_h) f^{\prime}(a) >0, \label{b2}
\end{equation}
Then, using the approximation \eqref{approximation1}, we obtain:
\begin{equation}
\Big[f^{\prime} (r_h)  \frac{e^{\kappa x}+1}{e^{\kappa x}-1}- f^{\prime \prime}(r_h) (a-r_h) \Big] (a-r_h) >0.
\end{equation}
Since the first derivative of $f(r)$ at the horizon is related to the surface gravity of black holes: $\kappa = \frac{f^{\prime}(r_h)}{2}$. It must be non-negative. While the second derivative of $f(r)$ is associated to the Ricci scalar curvature $R$ \eqref{Ricci}, and can be positive or negative.
\par When the island is beyond the horizon, $a-r_h >0$, the condition \eqref{b2} is reduced to:
\begin{equation}
\frac{f^{\prime} (r_h)}{(a-r_h)} \cdot \frac{e^{\kappa x}+1}{e^{\kappa x}-1} > f^{\prime \prime}(r_h), \label{b4}
\end{equation}
at large distances, $b \gg r_h$, we have:
\begin{equation}
\begin{split}
x=r_{\star}(b) - r_{\star}(a) &\simeq \frac{b-r_h}{a-r_h}, \\
\frac{e^{\kappa x}+1}{e^{\kappa x}-1} & \simeq 1. \label{b5}
\end{split}
\end{equation}
Therefore, \eqref{b4} is becomes to
\begin{equation}
\frac{f^{\prime} (r_h)}{a-r_h} > f^{\prime \prime} (r_h). \label{b6}
\end{equation}
The LHS of the equation is a large positive number with order ${\cal O} (G_N^{-1})$, and RHS is order ${\cal O}(G_N)$ and finite. So the above conditions are always true except in the special case of $f^{\prime}(r_h)=0$ (which corresponding to the extremal black hole).
\par The other case is that the island is inside the horizon, $a-r_h<0$. Similarly, we just need to flip the symbol:
\begin{equation}
\frac{f^{\prime} (r_h)}{r_h-a} < f^{\prime \prime} (r_h). \label{b7}
\end{equation}
Based on the previous discussion, we find that this inequality can not be reached except in the extremal case $(f^{\prime} (r_h)=0$. Accordingly, the island must always outside the horizon. Substituting $a \sim r_h$ into \eqref{b1}:
\begin{equation}
\begin{split}
y(r) & \simeq \frac{3 A^{\prime} (a) f^{\prime}(r_h) (a-r_h)^2 }{f^{\prime} (r_h) (a-r_h)  \frac{e^{\kappa x}+1}{e^{\kappa x}-1}-(a-r_h) f^{\prime} (r_h) } \\
& \simeq \frac{6 A^{\prime} (a) (a-r_h)}{ e^{\kappa r_{\star}(b)}  -\sqrt{\frac{a-r_h}{r_h}} -1} =cG_N. \label{b8}
\end{split}
\end{equation}
One finally obtains the location of the island \eqref{location}.

\newpage

\section{Tabulation} \label{appendixc}
\quad In this appendix, we give some results for several black holes based on the result \eqref{condition1}. One can refer to Table \ref{table} below:
\begin{table}[htb]
\begin{center}
\fontsize{12}{26}\selectfont
\setlength{\tabcolsep}{0.5mm}
\begin{tabular}{|c|c|c|c|c|}
\hline
Black Hole   &$f(r)$       &Area   &$f^{\prime \prime}(r)$   &Location of Islands          \\[10pt]\hline
CGHS         &$1-e^{-2 \lambda (r-r_h)}$     &$e^{2 \lambda r}$   &$-4 \lambda^2$           &$a\simeq r_h+ \frac{cG_N}{12\lambda e^{2 \lambda r_h}} e^{\lambda r_{\star}(b)}$                                            \\[10pt]\hline
JT           &$\frac{r^2 -r_h^2}{\ell^2}$    &$\frac{r}{\ell}$            &$\frac{2}{\ell^2}$        &$a\simeq r_h+ \frac{cG_N \ell}{6}e^{\frac{r_h}{\ell^2} r_{\star}(b)}  $      \\[10pt]\hline
BTZ          &$\frac{r^2 -r_h^2}{\ell^2}$    &$2 \pi r$                 &$\frac{2}{\ell^2}$           &$a\simeq r_h+ \frac{cG_N}{12\pi} e^{\frac{r_h}{\ell^2} r_{\star}(b)}$                \\[10pt]\hline
Sch.     &$1-\frac{r_h}{r}$           &$4 \pi r^2$                 &$-\frac{2}{r_h^2}$          &$a\simeq r_h+ \frac{cG_N}{48\pi r_h} e^{\frac{r_{\star}(b)}{2r_h}}$         \\[10pt] \hline
Sch. AdS &$1-\frac{r_0}{r}+\frac{r^2}{\ell^2}$               &$4\pi r^2$             &$\frac{2}{\ell^2}-\frac{r_0}{r_h^3}$             &$a\simeq r_h+ \frac{cG_N}{48 \pi r_h} e^{\big(\frac{r_h}{\ell^2}+\frac{r_0}{2r_h^2}\big) r_{\star}(b)}$                    \\[10pt] \hline
Sch. dS  &$\frac{(r-r_h)(r-r_U)(r_C-r)}{3r \ell^2} $                   &$4 \pi r_h^2$                  &$\frac{2(r_C r_U -r_h^2)}{3 \ell^2 r_h^2}$                                   &\tiny{$a\simeq r_h+\frac{cG_N e^{\frac{(r_c-r_h)(r_h-r_u)}{6r_h\ell^2}}}{48 \pi r_h e^{-r_{\star}(b)}} $}                                  \\[10pt] \hline
RN                &$\big(1-\frac{r_h}{r} \big) \big(1-\frac{r_-}{r} \big)$                   &$4 \pi r^2$             &$-\frac{2(r_h-2r_-)}{r_h^3}$                                   &$a\simeq r_h+\frac{cG_N}{48 \pi r_h} e^{\frac{r_h-r_-}{2r_h^2} r_{\star}(b)} $                   \\[10pt] \hline
\makecell[c]{Higher \\-dimensional Sch.}&$1-\big(\frac{r_h}{r}\big)^{d-3}$                       &\tiny{$\Omega_{d-2} r^{d-2}$}                 &$-\frac{6-5d+d^2}{r_h^2}$                                   &$a\simeq r_h +\frac{cG_N e^{\frac{d-3}{2r_h}} r_{\star}(b) }{12\Omega_{d-2} r_h^{d-3}} $                                           \\[10pt] \hline
\end{tabular}.
\caption{The location of islands in several black holes.}
\label{table}
\end{center}
\end{table}

\end{document}